\documentclass[a4paper,10pt]{article}

\usepackage[utf8x]{inputenc}

\usepackage[T1]{fontenc}
\usepackage{amsmath,amssymb,bm}
\usepackage{sidecap}
\usepackage{times}
\usepackage{braket}

\usepackage{graphicx,epsf}
\usepackage{float}
\usepackage[left=2.5cm,top=2.5cm,right=2.5cm,bottom=3cm,nohead,nofoot]{geometry}

\usepackage{bbold}
\usepackage{wrapfig}

\usepackage{natbib}

\usepackage{xcolor} 

\date{}              
\begin{document}
\begin{centering}
\LARGE\textbf{Spin-orbit interaction in core-shell semiconductor-metal nanowires}

\vspace{12pt}      
\normalsize\textbf{Tudor-Gabriel Dumitru$^{\dagger}$, Anna Sitek$^{\ast}$, Gunnar Thorgilsson$^{\dagger}$, Sigurdur I.\ Erlingsson$^{\dagger}$, and Andrei Manolescu$^{\dagger}$}

\vspace{0pt} 
\normalsize\textit{$^{\dagger}$Department of Engineering, Reykjavik University, Menntavegur 1, IS-102 Reykjavik, Iceland}\\
\normalsize\textit{$^{\ast}$Institute of Theoretical Physics,  Wroclaw University of Science and Technology, 50-370 Wroclaw, Poland}\\ 
\vspace{12pt}  
\end{centering}
\vspace{6pt}

\noindent
\textbf{ABSTRACT}
\vspace{3pt}

We study theoretically the spin-orbit interaction of electrons confined in a tubular semiconductor nanowire, between an inner semiconductor core and an outer metallic extra shell. A band off-offset potential is present at the inner semiconductor-semiconductor interface and a more complex potential barrier at the outer metal-semiconductor contact.  The cross section of the nanowire has a hexagonal geometry.  We use a model derived
from the $\bm{k}\!\cdot\!\bm{p}$ method, and discuss the effects of the interface potentials on the strength of the spin-orbit coupling and on the localization of the wave functions within the semiconductor shell \\


\vspace{12pt} 
\noindent
\textbf{1. INTRODUCTION}
\vspace{3pt}

Core-shell nanowires are prominent radial heterostructures widely utilized in quantum devices. Their various properties emerge from both the specific material pairing of the core and shell, as well as the high sensitivity of their electronic band structure to the cross-sectional geometry. Specifically, the lowest-energy states localize at the corners, and in the case of narrow shells lead to the formation of isolated quantum wires separated from the remaining spectrum by an energy gap. Conversely, the higher-energy states localized along the facets behave as coupled structures \cite{Sitek15,Sitek19} . 

Since a single nanowire can emulate coupled multi-wire systems, this makes it capable of hosting multiple Majorana zero modes \cite{Manolescu17, Stanescu18}. The stabilization of such topological phases and the broader functionality of spin-based quantum devices \cite{Nadj_Perge10} are based on the spin-orbit interaction (SOI) which arises either from bulk (Dresselhaus) or structural (Rashba) inversion asymmetry. The Rashba contribution is of particular interest, as it is explicitly determined by the local electrostatic confinement and because it is very sensitive to the spatial profile of the potential landscape, it can be engineered in various ways.  One such particular case is surrounding a semiconductor nanowire with a thin metallic layer that is also a superconductor \cite{Vaitiekenas20} leading to the creation and study of Majorana zero modes. However, initial studies have shown that the arising internal SOI coupling 
is very weak, not able to sustain such states \cite{Woods19}.

In the present work, we analyze how the Z-like potential profile due to an external metallic layer enhances the localization of the transverse wave functions and affects the overall SOI strength.  We utilize a model derived from the $\bm{k}\!\cdot\!\bm{p}$ formalism to explore how this penetrative interface potential modulates the Rashba-type SOI and the electronic structure within the semiconducting shell.
The remainder of the paper is organized into three parts: Section 2 details the theoretical model and the methodology used to obtain the metal-induced potential profile; Section 3 presents the results; and Section 4 summarizes the final conclusions.

\vspace{12pt} 
\noindent
\textbf{2. METHODS}
\vspace{3pt}

We study a structure consisting of an InP core surrounded by InAs and Al shells. We model
a full-shell metallic contact in the form of an external potential that penetrates 10 nm into the semiconductor shell. Computed via density functional theory (DFT) using the SIESTA software package \cite{Soler_2002}, this potential profile resembles the Z-shaped heterojunction characteristic of AlGaAs/GaAs systems, which have been widely used in electronics. 
Furthermore, we take into account the potential arising from the band offset between the core and semiconducting shell and assume intermixing of of InP and InAs atoms \cite{Sitek25b}.
We denote by $d$ the thickness of the InAs shell. Within this shell, the extended band-offset potential decays linearly from a maximum of 656.6 meV down to zero along the distance \(r\) measured from the InP-InAs interface. To incorporate the metallic contact, we introduce the parameter $s$ to quantify the inward penetration length of the potential due to the metallic shell, measured from the exterior boundary inwards. The two potential landscapes can overlap where their spatial domains intersect Figure 1(a).

The \(\bm{k}\!\cdot\!\bm{p}\) method and folding-down procedure \cite{Fabian07, Wojcik21} 
results in an infinite wire Hamiltonian 
\begin{equation}
\label{HSOI}
\mathcal{H}_{cc} =
\bigg[\frac{\hbar^2}{2m^{*}}\left(  k_{z}^2  
-\frac{\partial^2}{\partial x^2}  
-\frac{\partial^2}{\partial y^2} \right)  
 + V_{\mathrm{BO}}(x,y) + V_{\mathrm{MC}}(x,y)
\bigg] \sigma_0 + k_z \bigg(\alpha_x(x,y)\sigma_x + \alpha_y(x,y)\sigma_y\bigg) \ ,
\end{equation}
where  \(m^*\) is the effective mass and \(k_z\) is the wave vector along  the nanowire growth direction. The potential \(V_{\mathrm{BO}}(x,y)\) represents the extended band-offset potential \cite{Sitek25b}, while \(V_{\mathrm{MC}}(x,y)\) stands for the  potential due to the metallic contact, having a Z shape with a minimum of -50 meV and a maximum of 30 meV. This second interface potential was derived from first-principles density functional theory (DFT) calculations using the generalized gradient approximation (GGA) to evaluate the Hartree potential across the material junction. Exploiting the inherent translational symmetry of the crystal lattice, we extracted the bulk potential values from a representative supercell subjected to periodic boundary conditions. These bulk values subsequently serve as the basis for generating a dimensionless profile of the contact potential.

The metallic contact potential comprises two primary components: a dimensionless spatial profile and an averaged oscillation amplitude within the heterostructure, denoted as $A_{\mathrm{avg}}$. The dimensionless profile is defined by the Hartree potential of the heterostructure 
($V_{\mathrm{het}}(x,y)$) 
relative to the bulk value of InAs ($V_{\mathrm{InAs}}(x,y)$) and normalized by the energy difference between the two bulk materials ($V_{\mathrm{Al}}(x,y)-V_{\mathrm{InAs}}(x,y)$). Therefore, the effective metallic contact potential takes the form: 
\begin{equation}
    V_{\mathrm{MC}}(x,y) = A_{\mathrm{avg}} 
   \cdot 
%
       \left ( \frac{V_{\mathrm{het}}(x,y)-V_{\mathrm{InAs}}(x,y)}{V_{\mathrm{Al}}(x,y)-V_{\mathrm{InAs}}(x,y)} \right )
   \  .
\end{equation}

\begin{figure}[!ht]
	\centering
	\includegraphics[trim=0 0 0 10, clip, scale=0.25]{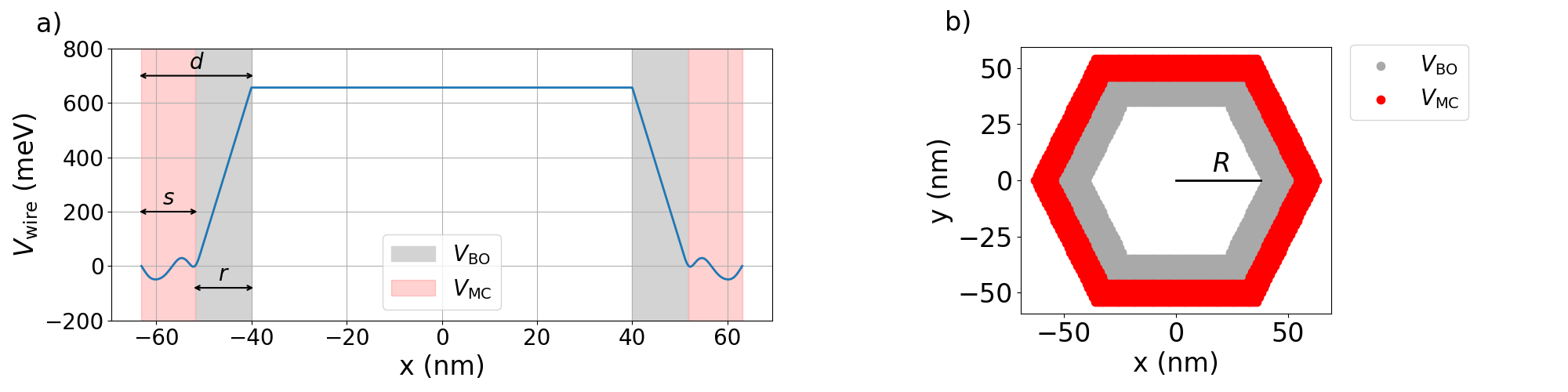} 
    \vspace{-20pt}
	\caption{Spatial profile of the confining potential across the nanowire. (a) 1D profile along the $x$-axis (corner to corner) detailing the potential magnitude. (b) The cross-sectional coverage of the potentials. This configuration assumes a shell thickness $d = 20$ nm. The band-offset decay length $r$ (gray region) and the external penetration depth $s$ (red region) are both set to 10 nm.}
	\label{fig:Potential}    
\end{figure}

The resulting Rashba SOI coefficients present in the Hamiltonian are obtained from the following set of equations
\begin{equation}
    \alpha_x(x,y) = -\alpha_0 \frac{\partial}{\partial y}V_\mathrm{wire}(x,y) 
    \qquad \text{and} \qquad
   \alpha_y(x,y) =  \alpha_0 \frac{\partial}{\partial x}V_\mathrm{wire}(x,y) 
   \ , 
\end{equation}
with \(\alpha_0\) representing the amplitude dependent on the InAs shell semiconductor band gap as well as the spin-split gap and the interband momentum matrix elements \cite{Fabian07}. \(\sigma_x\) and \(\sigma_y\) stand for the Pauli matrices and $V_\mathrm{wire}(x,y)= V_{\mathrm{MC}}(x,y)+V_{\mathrm{BO}}(x,y)$ represents the full potential inside the wire.

\vspace{12pt} 
\textbf{3. RESULTS} 
\vspace{3pt}

Below we present the results for a \(20\)-nm-thick InAs shell that surrounds an InP core of radius $R=40$ nm. We investigate the combined effects of the external Al layer and of the inner InP core on the SOI within the InAs shell.

\vspace{3pt}
\noindent
\textbf{3.1 States of the wire without internal potential}
\vspace{3pt}

\begin{figure}[!ht]
\centering
\includegraphics[trim=0 0 0 50, clip, scale=0.16]{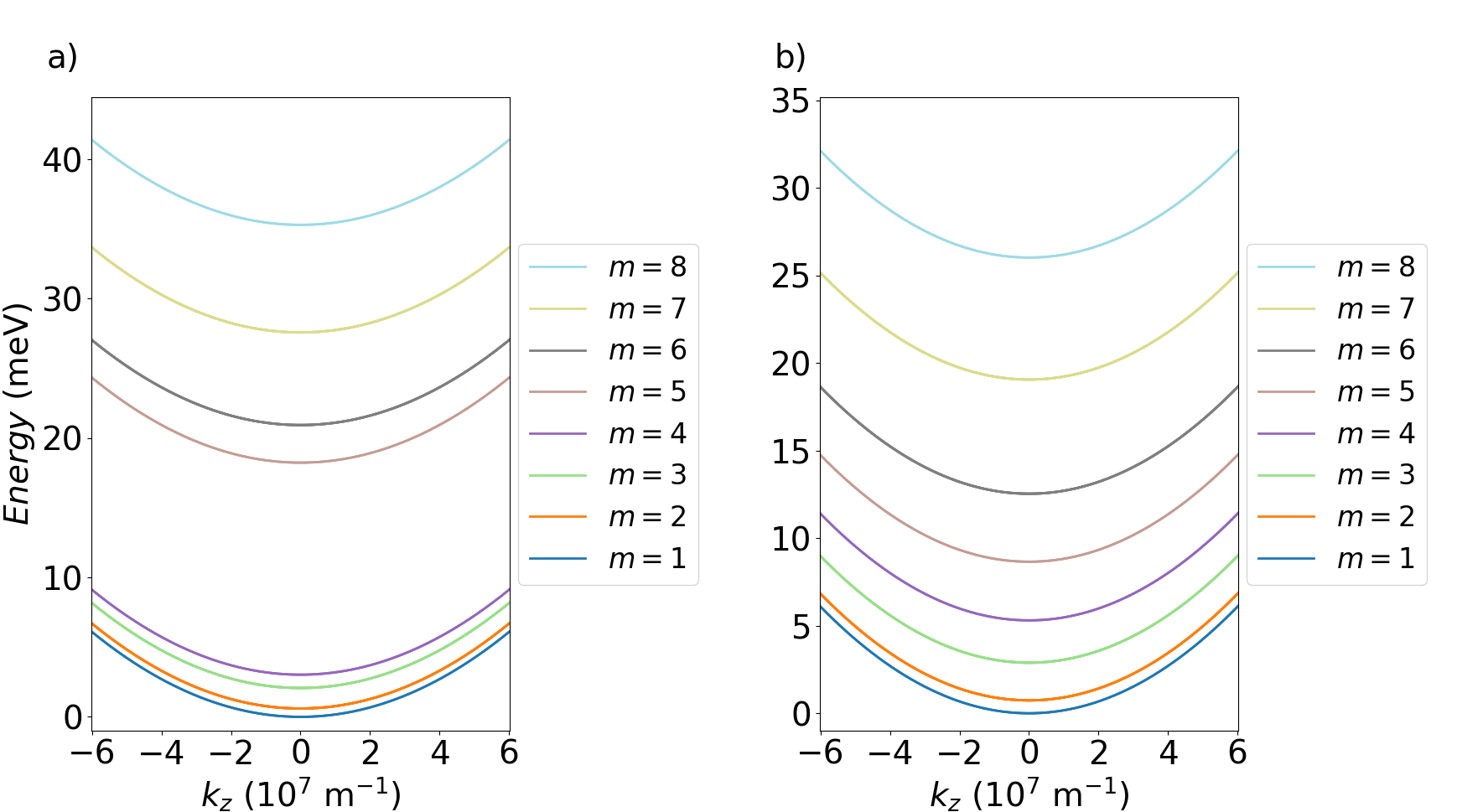}    
\caption{Energies of the first 24 states for a shell thickness of $d=10$ nm (a) and $d=20$ nm (b), using $V_{\mathrm {wire}}=0$.}
\label{fig:SOI_r}    
\vspace{-10 pt}
\end{figure}

The states of the wire obtained by diagonalizing the Hamiltonian  Eq.~\ref{HSOI} without the internal potential, i.e. $V_{\mathrm {wire}}(x,y)=0$, which also means without SOI, are grouped into sets of 12 parabolas (twice the number of corners). Within each of these sets, the lowest and highest (\(m=1, 4, 5, \text{ and } 8\)) are twofold degenerate whilst the middle ones are fourfold degenerate, Fig.2. The lowest 12 states are localized on the corners of the cross section whereas the higher states will have their maxima predominantly on the sides. 
Notably the large energy gap $\Delta$, which in Fig. 2(a) separates the corner-localized states decreases and for the \(20\)-nm-thick shell studied in this paper, it becomes comparable to other energy splittings. We focus on the thicker sample to match typical experimental setups.

\vspace{3pt}
\noindent
\textbf{3.2 SOI impact on the states}
\vspace{3pt}

The SOI discussed in this section lifts the fourfold degeneracies at finite wave vectors and shifts pairs of states in the $\pm k_z$ directions. The strength of this energy splitting is highly dependent on the potential profile, becoming significantly enhanced as the extended band offset potential penetrates deeper into the shell \cite{Sitek25a, Sitek25b}. The resulting energy spectrum exhibits twofold degenerate states.

As shown in Figure 3(a), the Rashba SOI induced solely by the short-range (\(r=10\) nm) extended band-offset potential is relatively weak. Introducing the metallic contact, Figure 3(b), enhances this coupling since the radial electric field and thus the non-zero Rashba coefficients distribution now span the entire cross-section of the shell. The combined potentials shift the electrons towards the metallic shell effectively narrowing the electron localization area which results in a large separation of corner-localized states from side-localized ones. However we still notice a small SOI effect. To generate a stronger SOI we would need to amplify the metallic potential three times, and the result is shown in Figure 3(c).

\begin{figure}[!ht]
	\centering
	\includegraphics[trim=0 0 0 50, clip, scale=0.16]{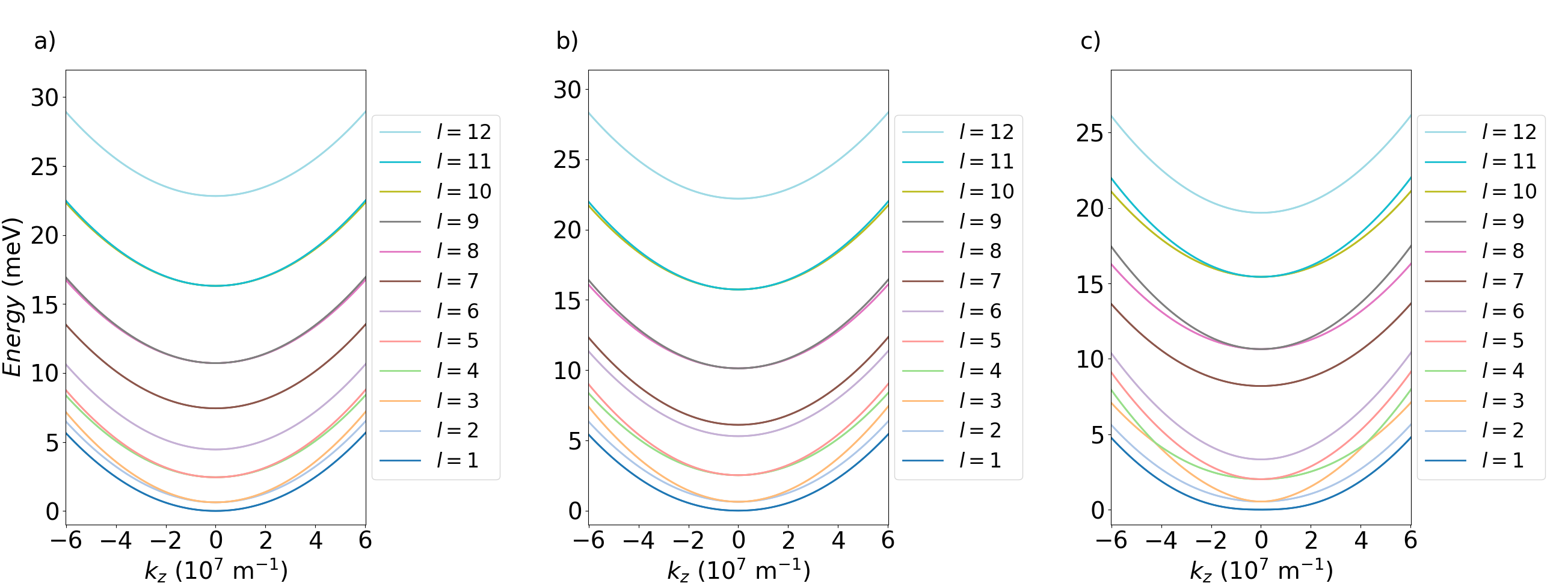}  
	\caption{Spin-orbit effect on the energy dispersion of the first 24 states for a shell thickness $d=20$ nm and penetration length $r=10$ nm. Panel (a) shows the case of zero external penetration ($s=0$) with the effects being attributed entirely to the extended band-offset potential, while (b) and (c) show the effects for a penetration length of $s=10$ nm. Panel (b) uses a well depth of $-50$ meV and a barrier height of $30$ meV, whereas (c) uses $-150$ meV  for the depth of the well and a $90$ meV barrier.}
	\label{fig:BandsSOI10}    
\end{figure}

\begin{figure}[!ht]
	\centering
	\includegraphics[trim=0 0 0 50, clip, scale=0.16]{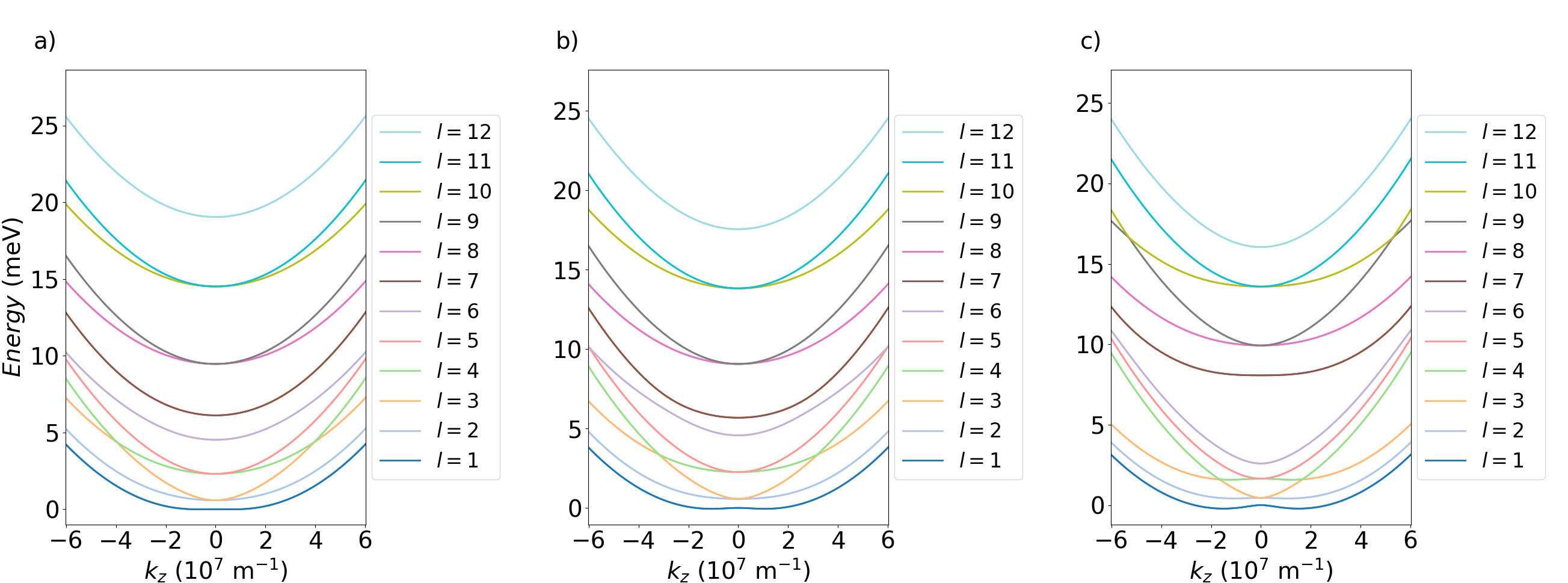}  
\vspace{-3pt}   
	\caption{Spin-orbit effect on the energy dispersion of the first 24 states for a shell thickness $d=20$ nm and penetration length $d=r=20$ nm. Panel (a) shows the case of zero external penetration ($s=0$) with the effects being attributed entirely to the extended band-offset potential, while (b) and (c) show the effects for a penetration length of $s=10$ nm. Panel (b) uses a well depth of $-50$ meV and a barrier height of $30$ meV, whereas (c) uses $-150$ meV  for the depth of the well and a $90$ meV barrier.
\vspace{-9pt}    
    }
	\label{fig:BandsSOI20}    
\end{figure}

For the second case, the overall effects due to SOI are notably stronger, Fig. 4, though this enhancement stems from the band-offset potential spanning the entire shell thickness rather than the metallic contact itself \cite{Sitek25a, Sitek25b}. When the metallic contact $V_{\mathrm{MC}}(x,y)$ is added and superimposed on the extended band-offset potential (Figs. 4b and 4c), its additional impact on the Rashba potency is minimal; instead, its primary consequence remains the enhanced spatial localization of the corner and side states.

\vspace{12pt} 
\noindent
\textbf{4 Conclusions}
\vspace{3pt}

We studied a hexagonal a core-shell structure consisting of a semiconducting core and two shells, one semiconducting and the second metallic.
We analyzed the effects of an extended band-offset potential at the core-semiconductor shell heterojunction and a Z-like potential induced by the external metallic shell on the Rashba SOI within a semiconducting shell. We show that in the presence of both potentials only the fourfold degeneracies are lifted resulting in twofold degenerate states at finite wave vectors. The potential due to the metallic contact only slightly increases the SOI, which is in line with a previous study \cite{Woods19}. Instead, the dominant effect is the effective narrowing of the electron distribution area, and thus widening the energy separation of the corner-localized states. In the case of narrow structures the SOI leads to the formation of six camelback-shaped energy states in each group of twelve states.

\vspace{12pt}
\noindent
\textbf{ACKNOWLEDGEMENTS}\\
\noindent
This work was financed by Reykjavik University Research Fund, Grant 223016. We are very thankful to George Alexandru Nemnes and Nicolae Filipoiu for their assistance with the atomistic calculations.

\noindent

\setlength{\bibsep}{0pt plus 0.3ex}
\bibliographystyle{unsrt}
\bibliography{core_shell.bib}

\end{document}